# Polarization Pinning at Antiphase Boundaries in Multiferroic YbFeO$_3$

Guodong Ren[1], Pravan Omprakash[1], Xin Li[2], Yu Yun[2,3], Arashdeep S. Thind[1], Xiaoshan Xu[2,4], Rohan Mishra[5,1*]

[1]*Institute of Materials Science and Engineering, Washington University in St. Louis, St. Louis, MO 63130, USA*

[2]*Department of Physics and Astronomy, University of Nebraska, Lincoln, NE 68588, USA*

[3]*Department of Mechanical Engineering & Mechanics, Drexel University, Philadelphia, PA 19104-2875, USA*

[4]*Nebraska Center for Materials and Nanoscience, University of Nebraska, Lincoln, NE 68588, USA*

[5]*Department of Mechanical Engineering and Material Science, Washington University in St. Louis, St. Louis, MO 63130, USA*

*\*Email: [rmishra@wustl.edu](mailto:rmishra@wustl.edu)*

**ABSTRACT**

The switching characteristics of ferroelectrics and multiferroics are influenced by the interaction of topological defects with domain-walls. We report on the pinning of polarization due to antiphase boundaries in thin films of the multiferroic hexagonal YbFeO$_3$. We have directly resolved the atomic structure of a sharp antiphase boundary (APB) in YbFeO$_3$ thin films using a combination of aberration-corrected scanning transmission electron microscopy (STEM) and total energy calculations based on density-functional theory (DFT). We find the presence of a layer of FeO$_6$ octahedra at the APB that bridge the adjacent domains. STEM imaging shows a reversal in the direction of polarization on moving across the APB, which DFT calculations confirm is structural in nature as the polarization reversal reduces the distortion of the FeO$_6$ octahedral layer at the APB. Such APBs in hexagonal perovskites are expected to serve as domain-wall pinning sites and hinder ferroelectric switching of the domains.

## I. INTRODUCTION

Topological defects, such as dislocations and domain boundaries, can alter the properties of their host materials. For instance, dislocation cores in an insulator can be metallic and serve as conducting nanowires.[1,2] In ferroelectrics, boundaries between adjacent domains having different directions of polarization can have enhanced local conduction.[3,4] These domain walls can act as sinks and reservoirs for point defects and lead to large gradients in strain and electric field.[5] They intricately affect domain switching,[6] and determine the response of ferroelectrics.[7] The large stress fields around topological defects have been used to promote the formation of metastable phases — that are otherwise challenging to grow using other techniques.[8] Furthermore, topological defects are dynamic as they can appear, grow and disappear in response to electric fields or stress. Thus, they have been proposed as dynamically configurable electronic components.[9]

The properties of topological defects are determined by their local structure and composition. Aberration-corrected scanning transmission electron microscopy (STEM) has become a powerful tool for resolving the structure of such defects. The simultaneous use of multiple imaging detectors allows the direct determination of the position of atomic columns belonging to both heavy and light elements.[10] Energy dispersive X-ray spectroscopy (EDS) and electron energy-loss spectroscopy (EELS) carried within STEM enable chemical composition analysis with atomic-resolution. STEM-EELS also provides information about the electronic structure and local bonding. Dr. Steve Pennycook — to whom this special issue is dedicated to — has been at the forefront of developing STEM-EELS,[11-15] and using them to elucidate the structure of topological defects in various materials.[16,17] Furthermore, by collaborating with experts in atomic scale modeling, such as with the group of Prof. Sokrates Pantelides at Vanderbilt University, Pennycook and co-workers have established the combination of STEM-EELS with atomic scale modeling and electronic structure calculations as an invaluable set of tools to resolve the structure of topological defects and understand their properties.[18-22] STEM provides a two-dimensional (2D) projection of the 3D atomic structure. Total-energy calculations based on density-functional theory (DFT) provide a way to compare different 3D models and identify the structure having a reasonable energy under the growth conditions and the best match with the acquired information from STEM-EELS. Moreover, by providing access to the electronic structure of the defective material, DFT calculations allow for a direct comparison of the physical properties in the presence of defects with those measured experimentally. Thus, STEM-EELS along with DFT have now become a mainstream method for probing topological defects.[23]

In this article, we have utilized a combination of first-principles DFT calculations with aberration corrected STEM to directly resolve the effect of antiphase boundaries on the polarization in a multiferroic material h-YbFeO$_3$ (h stands for hexagonal). Rare-earth manganites and ferrites with the hexagonal structure show simultaneous ferroelectric and magnetic ordering.[24-30] Some of them also show strong magnetoelectric coupling and have been used to demonstrate electric-field switching of magnetization.[25,31,32] Recently, there has been a growing focus on the role of defects of various dimensionalities on the magnetic and ferroelectric behavior of these hexagonal multiferroics. Barrozo *et al*. reported a reduction in the domain-wall migration energy in h-LuFeO$_3$ with increasing concentration of oxygen vacancies, which was controlled by changing the oxygen partial pressure during synthesis.[33] They observed a 40% reduction in the switching voltage for films having a large concentration of oxygen vacancies. Skjærvø *et al*. reported an increase in *p*-type electronic conductivity in h-YMnO$_3$ samples annealed in oxygen-rich atmospheres.[34] They used DFT calculations to attribute the observed increase in hole conductivity to the presence of highly mobile oxygen interstitials in the bulk. Evans *et al*.[35] demonstrated the creation of partial dislocations in h-Er(Mn,Ti)O$_3$ single crystals using the local electric field from the tip of a scanning probe. They used STEM imaging to characterize the electric-field-injected dislocations and stacking faults associated with swapping of Er and Mn atomic columns. Higher dimensional planar defects, for instance stacking faults and APBs with complex structures,[36-38] have also been reported in hexagonal multiferroics and attributed to the breaking of the ferroelectric vortices.

Multiferroic h-*RM*O$_3$ (*R* = rare earth ion, *M* = Mn, Fe) show improper ferroelectricity, where the polar order arises as a secondary effect, driven by a primary-order parameter that is often another structural distortion.[26,30] While the role of defects in polarization switching of proper ferroelectrics is generally well established, such understanding for improper ferroelectrics including multiferroic h-*RM*O$_3$ remains an ongoing area of investigation. Given the limited understanding of high dimensional planar defects in improper ferroelectrics, we have initiated a study of APBs and their effect on polarization in multiferroic h-YbFeO$_3$. Here, we reveal the atomic structure of an APB within the multiferroic h-YbFeO$_3$ by synergistically combining STEM with DFT calculations. We find that ferroelectric domains with opposite polarization are pinned at the APB due to the geometric constraints imposed by a layer of FeO$_6$ octahedra. Additionally, the presence of a layer of FeO$_6$ octahedra at the APB can result in the accumulation of negative charges, which are likely to attract positively charged point defects. This study provides insights into polarization pinning due to topological defects such as APBs in hexagonal multiferroics.

## II. METHODS
### A. Crystal Growth

A h-YbFeO$_3$ thin film (40 nm thick) was grown on a CoFe$_2$O$_4$/La$_{2/3}$Sr$_{1/3}$MnO$_3$/SrTiO$_3$ (111) substrate by pulsed laser deposition (PLD) system with a KrF excimer laser (248 nm and 2 Hz repetition rate), at growth temperatures between 650 to 850 °C and oxygen pressure of 10 mTorr. Before the thin-film deposition, substrates were pre-annealed at 700 °C for 1 h. The $La_{2/3}Sr_{1/3}MnO_3$ layer (~ 30 nm) was grown at a substrate temperature of 700 °C and oxygen pressure of 80 mTorr on the $SrTiO_3$ (STO) substrate. The $CoFe_2O_4$ (CFO) layer (~ 10 nm) was grown at the temperature of 600 °C and the oxygen pressure of 10 mTorr. More details about the growth can be found elsewhere.[32,39]

## B. STEM characterization

An [001]-oriented specimen was prepared using Ar-ion milling (Fischione Model 1010). To enable the direct observation of $YbFeO_3$ structure along the [001] direction, a plan view specimen was first mechanically polished to remove the majority of the substrate underneath $YbFeO_3$. This was followed by a single-side thinning process utilizing a 4 keV ion beam with an angle of incidence at 5° to get rid of the $CoFe_2O_4/La_{2/3}Sr_{1/3}MnO_3/SrTiO_3$ (111) substrate. Another [100]-oriented lamella was prepared using a ThermoFisher Scios 2 DualBeam FIB. We performed STEM imaging using an aberration-corrected Nion UltraSTEM 100 operated at 100 kV with a convergence semi-angle of 30 mrad. High-angle annular dark-field (HAADF) images were acquired using an annular dark-field detector with inner and outer collection semi-angles of 80 and 200 mrad, respectively. We recorded the HAADF data as a sequence of 20 images acquired during a fast scan (dwell time of 1 μs), which were later aligned by correcting for sample drift. Electron energy loss spectroscopy (EELS) was carried out using a Gatan Enfina EEL spectrometer attached to the Nion UltraSTEM. A collection semi-angle of 48 mrad and an energy dispersion of 1 eV per channel was used to acquire EELS data. Quantitative EELS analysis was performed using DigitalMicrograph.

STEM-HAADF simulations were performed using the multi-slice method as implemented in μSTEM.[40] The sample thickness was set to 15 nm and the defocus value was set to 10 Å to obtain good agreement in intensity profiles with the experimental data. We performed the image simulations using an aberration-free probe with an accelerating voltage of 100 kV and a convergence semi-angle of 30 mrad. The inner and outer collection angles for the HAADF detector were set to 80 and 200 mrad, respectively.

## C. DFT calculations

DFT calculations were carried out using projector augmented-wave potentials [41] as implemented in the Vienna Ab initio Simulation Package (VASP).[42] The PBEsol functional was used to describe the exchange-correlation interactions.[43] A plane-wave basis set with an energy cutoff of 500 eV and an energy convergence criterion of $10^{-6}$ eV for the electronic convergence were applied. A $k$-point spacing of 0.05 $Å^{-1}$ was chosen for Brillouin zone sampling. The structures were optimized until all forces on atoms were less than 0.01 eV/Å. To better describe the localization of Fe-$d$ states, we used PBEsol+$U$ with an effective on-site Hubbard $U = 5.0$ eV used for the Fe-$d$ electrons.[44] We considered different magnetic configurations for the calculations. The APB in YbFeO$_3$ was modeled using a 1×4×1 supercell of the orthogonal lattice of the $P6_3cm$ phase consisting of 220 atoms. The supercell has two APBs (to maintain the periodic boundary conditions) with different stoichiometries based on the STEM observations. For the non-shift configuration, each APB contained 4 Fe atoms and 8 O atoms, while for the shift configuration, the APB consisted of 4 Fe atoms and 6 O atoms. The supercells with different configurations were relaxed and compared to determine the optimal match with the experimental results.

## III. RESULTS AND DISCUSSION

As a representative of the multiferroic h-$R$MO$_3$ family, h-YbFeO$_3$ thin films stand out as a promising candidate for electrical-field driven modulation of magnetization, especially due to its enhanced domain-wall magnetoelectric coupling.[25,26,32] Below the Curie temperature, $T_C \approx$ 1000 K,[25,26] h-YbFeO$_3$ displays a non-centrosymmetric $P6_3cm$ symmetry, as depicted in Fig. 1a. This structure consists of corner-sharing FeO$_5$ trigonal bipyramids separated by layers of Yb$^{3+}$ ions, each of which are coordinated by eight oxygen atoms. Unlike the intensively studied proper ferroelectrics, such as BaTiO$_3$ and PbTiO$_3$, whose spontaneous polarization is the primary order parameter emerging from the ferroelectric transition, h-YbFeO$_3$ exhibits improper ferroelectricity arising from a geometric effect that couples the polar distortion to a nonpolar structural distortion.[26] The polar order in h-YbFeO$_3$, manifested as a rumpling of the Yb layer along the $c$-axis, is driven by a primary tilt trimerization of the FeO$_5$ bipyramids, which corresponds to the condensation of a zone-boundary K$_3$ mode of the high-symmetry $P6_3/mmc$ structure.[26,45-47] The primary K$_3$ mode distortion, characterized by a collective tilt of FeO$_5$ bipyramids, can be described by a two-component order parameter ($Q$, $\varphi$) having an amplitude $Q$ and a phase $\varphi$.[48-50] Generally, the displacement of Yb cations — that is related to the spontaneous polarization $P_s$ — can be described by a sinusoidal corrugation pattern:

$$u(x) = Q cos(\varphi - \boldsymbol{q} \cdot x), \tag{1}$$

where $u(x)$ is the displacement of the rare earth cations along $c$-axis at position $x$ and $\boldsymbol{q}$ is the wavevector of the K$_3$ mode. Because of the hexagonal symmetry, there are six distinct trimerized states, corresponding to six degenerate values of the phase: 0, ±π/3, ±2π/3 and π. This leads to the formation of the topologically protected ferroelectric domain-vortex patterns that are observed in most h-$R$MO$_3$ systems.[51-54]

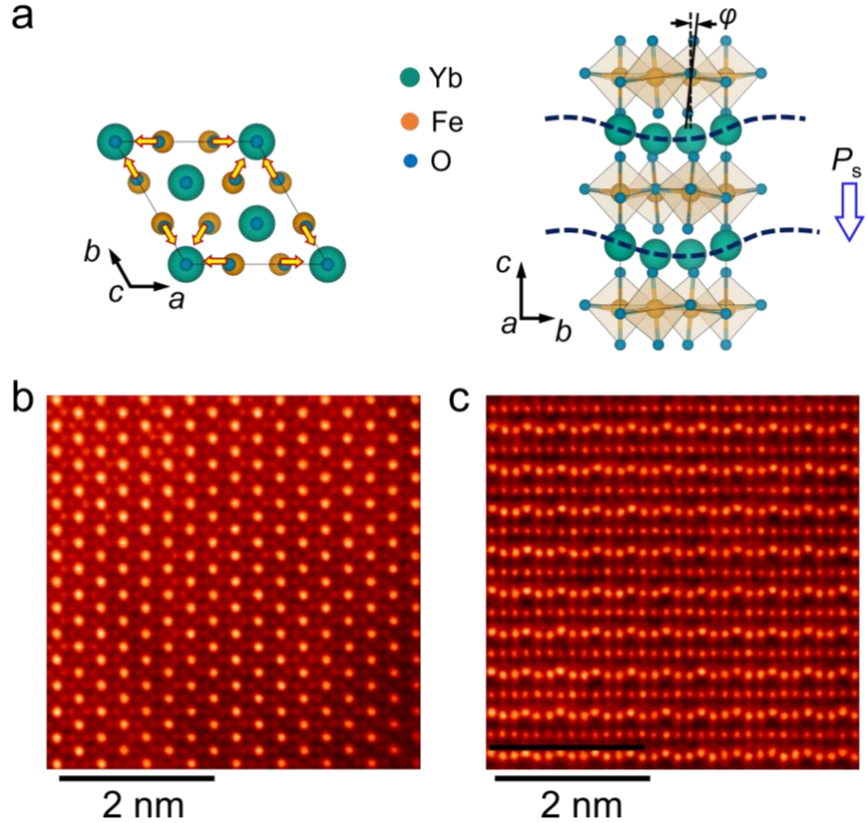

**Figure 1. a,** Schematics of h-YbFeO$_3$ structure viewed along [001] on the left and [100] on the right. The arrows in yellow represent the direction of displacement of the apical oxygen ion associated with the FeO$_5$ trigonal bipyramids and caused by the K$_3$ trimerization mode. The corrugated displacement of the Yb layers is induced by the collective tilt of the FeO$_5$ bipyramid to an in-plane rotation angle of $\varphi$. **b,** Atomic-resolution HAADF-STEM image viewed along the [001] direction with the hexagonal Yb sublattice appearing with the brightest intensity. **c,** Representative HAADF-STEM image taken along the [100] zone axis showing the polar displacement of Yb in a corrugated pattern. Scale bars are 2 nm in (**b**-**c**).

To resolve the polar order in h-YbFeO$_3$, we carried out real-space imaging using HAADF-STEM with atomic resolution. The HAADF-STEM images were taken from [001] and [100] zone axes and are shown in Figs. 1(b, c), respectively. The intensity of atomic columns in a HAADF image is proportional to the square of the effective atomic number of columns ($Z^2$).[13] In the HAADF image of h-YbFeO$_3$ taken along the [001] zone axis in Fig. 1(b), the hexagonal sublattice of Yb ($Z = 70$) appears brightest, while the Fe atomic columns ($Z = 26$) are less bright. The O columns are invisible due to limitations in the dynamic range of the HAADF detector. In the HAADF image along the [100] zone axis in Fig. 1c, the polar displacement of Yb layers is readily recognizable from its characteristic $Q$-related corrugation displacement pattern. By fitting the Yb displacement with equation (1), we can identify that the region shown in Fig. 1c has a downward polarization throughout the image. We note that in these small regions shown in Figs. 1(b, c), the sample exhibits good crystallinity without any apparent crystallographic defects.

We also observed several defective nanoregions in the h-YbFeO$_3$ film. One such region having a sharp boundary is shown in the HAADF image in Fig. 2a that was taken along the [100] orientation. The defective region features a lattice disruption with a quarter-unit-cell (½ Yb layer) shift of registry along the *c*-axis with respect to adjacent region. This observation confirms the formation of an APB. We find the presence of a column of interstitial Fe atoms at the APB, which appear to be bridging the Fe layers on one side of the APB to the Yb layer on the other side, in this projection. To map out the atomic positions and the polarization around the APB, we fit the atomic position of Yb columns with a 2D-Gaussian. The displacement pattern of Yb layer was then fitted to equation (1) with $P_s \propto Q^3$, where the up- or down- orientation of the ferroelectric polarization is determined by the phase angle $\varphi$ (up: 0, $\pm 2\pi/3$; down: $\pm \pi/3$, $\pi$). From the polarization map superimposed on the HAADF, as shown in Fig. 2b, we observe a 180° inversion in the polarization across the entire APB. Polarization inversion across the APBs was also observed in other regions of the film, as shown in Fig. S2 (*Section 1 in Supporting Information*). The coincidence of the 180° domain wall with the APB hints at the potential pinning of polarization at the APB, a phenomenon that has been observed in other ferroelectric systems.[6,55,56] It is also important to note that, away from the APB, there is a polarization domain wall on the right side of Figure 2b, formed by structural variants involving the tilt of one of six degenerate FeO$_5$ bipyramids tilts. This phenomenon has been extensively studied in the literature.[47-54]

To identify any composition variation across the APB and the adjacent domains, atomically resolved EEL spectrum maps were acquired from a region delineated by the white dashed box in Fig. 2a. Elemental maps of Yb-$N_{4,5}$, Fe-$N_{2,3}$, and O-*K* edges, integrated after background subtraction, are shown in Figs. 2(c-e). It is evident that the Yb layers display a discontinuity across the APB, while additional Fe columns are located at the APB, which is consistent with observations from the HAADF image in Fig. 2a.

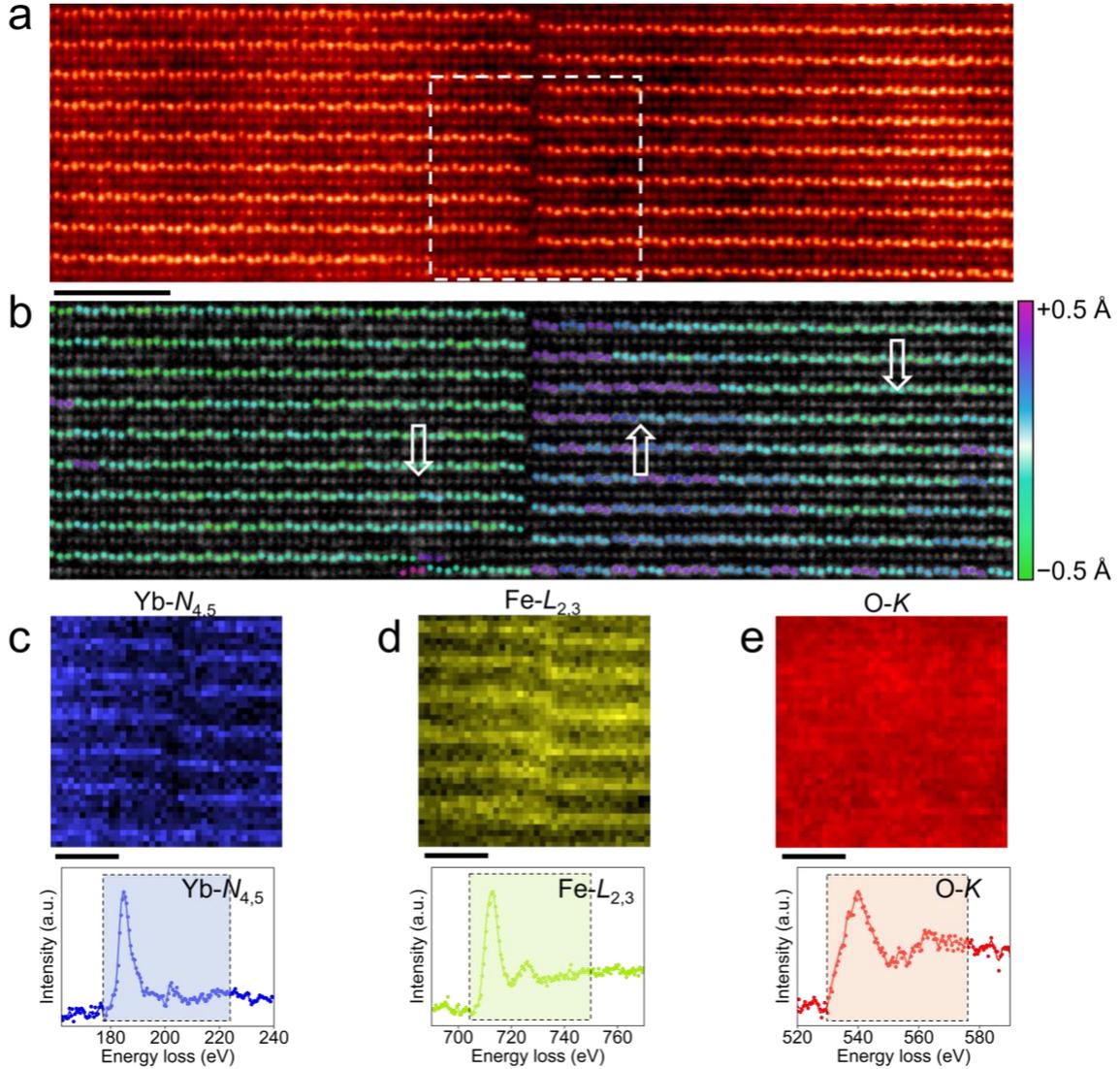

**Figure 2. Atomic-resolution STEM HAADF imaging and core-lose EELS mapping of APB in h-YbFeO$_3$ oriented along [100] axis. a,** Representative HAADF-STEM image of as-grown h-YbFeO$_3$ with an atomically sharp APB. **b,** Ferroelectric domain mapping across the APB in **a**, with the colored overlay indicating the polar displacement vector $Q\cos(3\varphi)$. The visualization reveals the polarization inversion across the APB. **c-e,** False-color elemental maps of Yb-$N_{4,5}$, Fe-$N_{2,3}$, and O-$K$ edges from the region in **a** highlighted with a white dashed box (top panel). The EEL spectra of corresponding elements used to make the elemental maps in (**c-e**) after PCA denoising and background subtraction (bottom panel). The energy range used for signal integration for each element is highlighted with colors. Scale bars represent 2 nm in (**a**–**b**) and 1 nm in (**c-e**).

To resolve the 3D atomic structure of the APB and investigate its interaction with the polarization of the adjacent domains, we modeled the defective structure using DFT. Following the STEM observations, we considered three different configurations of the APB with the additional $FeO_6$ layer, which are illustrated in Figs. 3(a-c). Specifically, model 1 shown in Fig 3a features extra Fe and O atoms forming $FeO_6$ polyhedra at the APB. In this model, opposite polarizations are initialized across the APB, with Yb atoms in dark green representing downward polarization and Yb atoms in pale yellow indicating upward polarization. These two regions with opposite polarization are aligned with a relative half-unit-cell shift along the *ab*-plane. In contrast, Model 2, shown in Fig. 3b, is constructed without shifting the regions with opposite polarization along the *ab*-plane. Furthermore, to examine ferroelectric domain switching across the APB, we also constructed a supercell with the same polarization orientation in the two adjacent domains, as shown in Fig. 3c (shifted single domain). The three supercells were then optimized for both volume and ionic positions until the forces acting on all atoms were below 0.01 eV Å$^{-1}$. Both the ferromagnetic (FM) and anti-ferromagnetic (AFM) configurations were relaxed, and their energies were compared. Both the FM configurations were higher in energy than their AFM counterparts (energy difference of 0.264 eV/f.u. for model 1 and 0.337 eV/f.u. for model 2). The comparison between the structure for both model 1 and model 2, before and after DFT optimization, can be found in Fig. S3 (*Supporting Information section II*).

The displacement pattern of the Yb layers — that is related to polarization reversal across the APB in Fig 3a, aligns most closely with the STEM observations. In contrast, the supercell in Fig. 3b, initially constructed with non-shifted domains, did not stabilize the initial polar displacement pattern after structural optimization. Specifically, the downward displacement pattern of the Yb layers (shown in dark green) underwent a $\pi/3$ phase shift near the APB during the structure optimization. The supercell model 3 in Fig. 3c, was found to be energetically unfavorable and was found to subsequently relax to the configuration with opposite polarizations across the APB, as depicted in Fig. 3a. Therefore, the structural optimization using DFT calculations support our hypothesis of the APB acting as a pinning site for the ferroelectric domains. This is because, in the supercell shown in Fig. 3a, the Fe located at the APB is bonded to oxygen atoms of $FeO_5$ polyhedra from adjacent domains across the APB resulting in the formation of $FeO_6$ polyhedra. Since the ferroelectric switching in h-YbFeO$_3$ is driven by the collective tilt of the $FeO_5$ trigonal bipyramids, the presence of the $FeO_6$ polyhedra at APB imposes geometric constraints on the $FeO_5$ tilting in the adjacent domains. To support adjacent domains with the same orientation of polarization, the $FeO_6$ octahedra at the APB would have to undergo significant distortions, resulting in the destabilization of the configuration in Fig. 3c. Consequently, this geometrical constraint contributes to the pinning of ferroelectric DW.

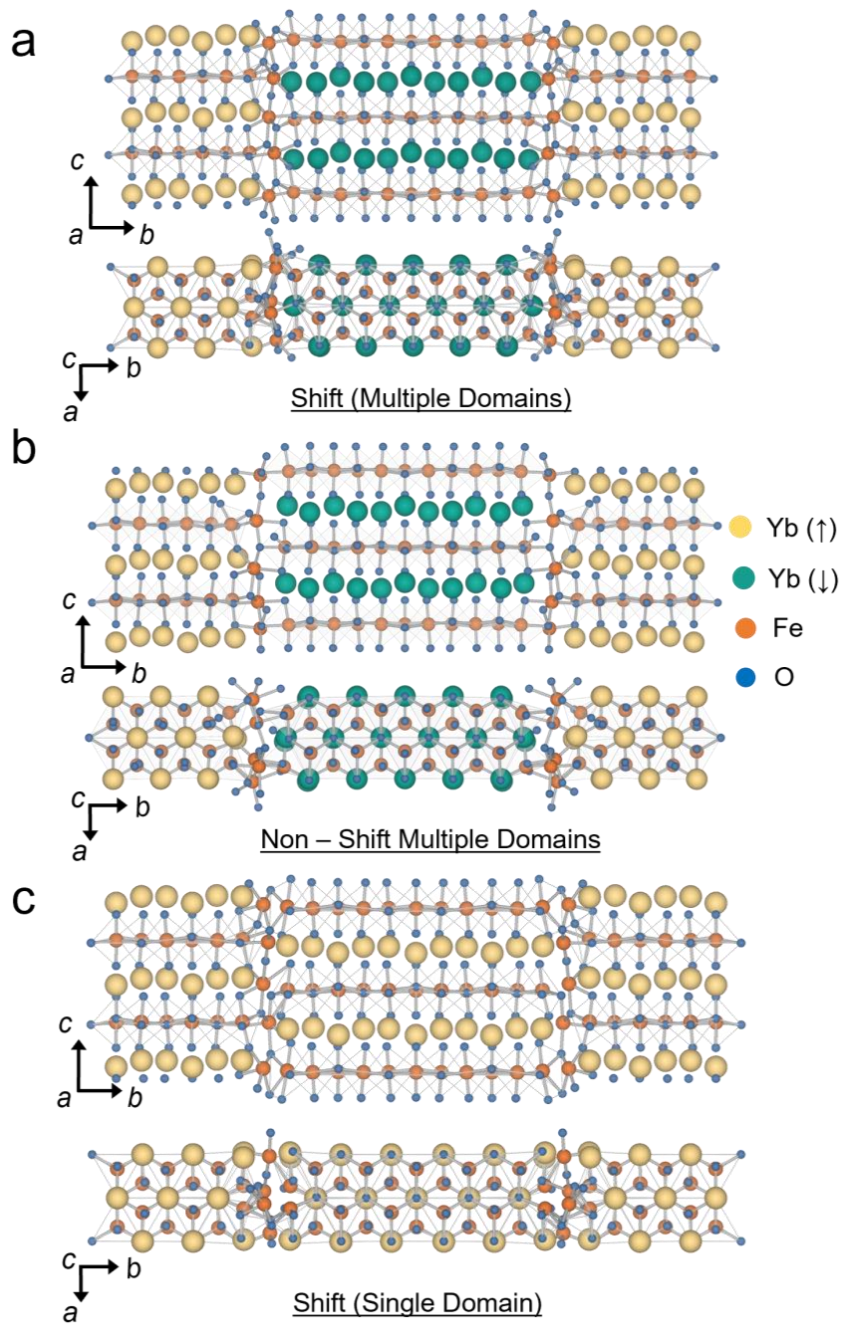

**Figure 3. Competing structural models of the APB. a,** DFT-optimized supercell with a half unit-cell shift along the *ab*-plane. The displacement patterns of Yb layers are opposite across the APB (Yb in dark green represents the downward displacements, while Yb in pale yellow corresponds to upward displacements). **b,** DFT-optimized supercell without a half unit-cell shift along the *ab*-plane. **c,** Supercell having a half unit-cell shift along the *ab*-plane but with the same polar displacement pattern across the APB before DFT optimization.

We further simulated the HAADF image of the APB model in Fig. 3a and compared it to the experimental image. We find an excellent agreement, as shown in Fig. 4(a, b). This is confirmed by a comparison of the intensity and spacing between the atomic columns along the APB, which is highlighted with a white dashed box in the experimental and the simulated HAADF images, as shown in Fig. 4c. We note that the horizontal "Fe layer" observed in the STEM image actually corresponds to superimposed oxygen and iron atomic columns. As depicted in the structural model in Figure 3, the oxygen atoms display a corrugation pattern along the horizontal direction due to the collective tilt of the $FeO_5$ bipyramids, which is responsible for the emergence of polarization. This corrugation pattern is visible in both the simulated and experimental HAADF-STEM images as intensity variation of the superimposed Fe/O atomic columns.

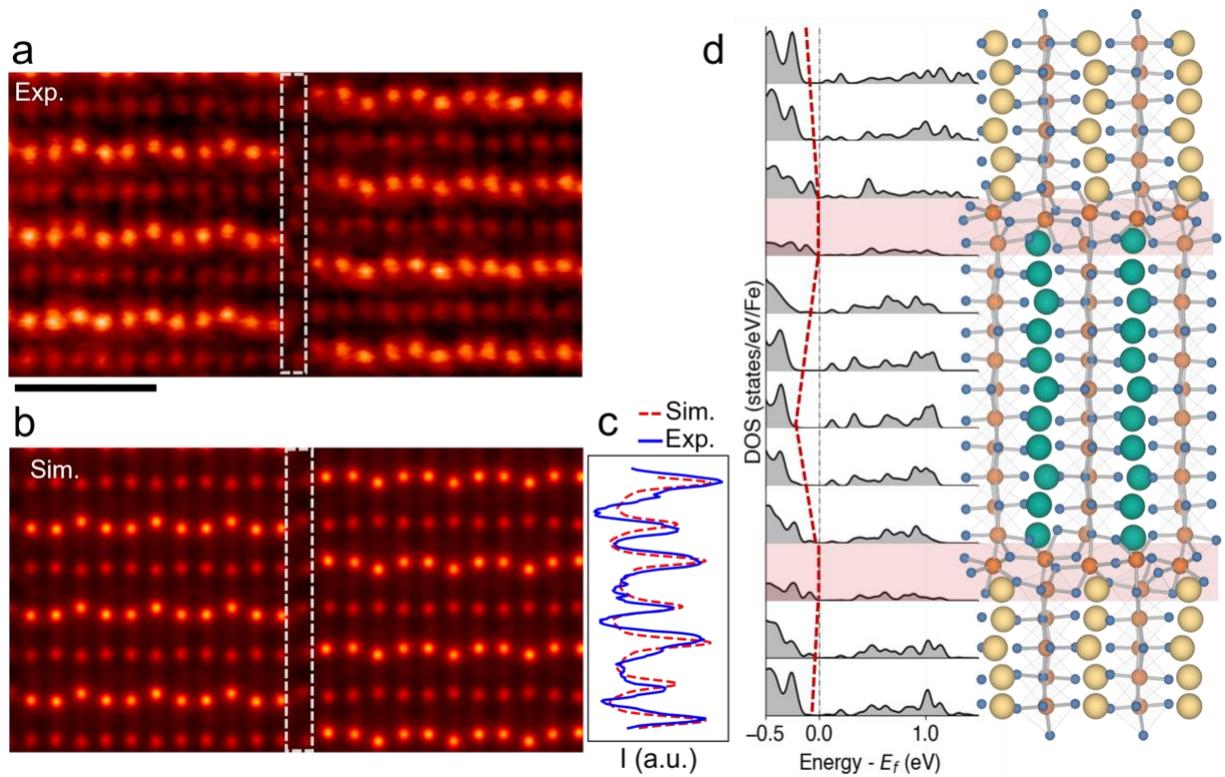

**Figure 4. Atomic and electronic structure of the APB.** Experimental HAADF image of the APB in **a** and the simulated HAADF image in **b** using the APB model shown in **d** (right panel). Scale bar represents 1 nm. **c,** Line profiles across the experimental and simulated HAADF images (highlighted using white boxes in **a** and **b**) comparing the intensity variation for the $FeO_6$ layer at the APB. d, (left panel) Calculated layer density of states (DOS) projected onto atomic planes across the supercell shown in d (right panel). The corresponding Fermi level for APB is marked by dashed gray lines.

Having resolved the atomic structure of the APB, we next calculated its electronic structure. The layer-projected density of states for the superlattice (Fig 4d) shows a shift of the valence states to higher energies upon approaching the APB from the centers of either domain. Specifically, the valence band maxima shifts to higher energies by ~160 meV upon approaching the APB, as indicated by the dashed red line in Fig. 4d. This band offset can be qualitatively correlated to the composition of the APB region, which includes an additional Fe-O layer. The higher energy states near the Fermi energy at the APB region come from the 3$d$ electrons introduced by the additional Fe atoms at the APB. Besides increasing the local *n*-type conductivity at the APB, the band bending and the electronic structure changes are also expected to attract holes and positively charged mobile ionic defects, such as oxygen interstitials that have been reported be abundant in hexagonal perovskites.[34] In the future, monochromated STEM-EELS measurements across the APB, may be used to validate the computationally predicted electronic structure changes.

## IV. CONCLUSION

In this work, we have resolved the atomic structure of an APB in h-YbFeO$_3$, and show its interaction with the polarization in the adjacent ferroelectric domains. STEM results reveal the formation of an atomically sharp APB separating the domains with 180° polarization inversion in the h-YbFeO$_3$ film. Elemental mapping using EELS and simultaneous HAADF-STEM observations suggest the presence of an additional Fe-rich vertical layer at the APB. DFT calculations suggest that the Fe-rich layer involves the formation of a layer of FeO$_6$ octahedra through the binding of the Fe atoms with the oxygen atoms from the adjacent domains. The geometric constraint imposed by the bridging FeO$_6$ octahedra at the APB, makes ferroelectric switching across the APB energetically unfavorable, as it would lead to significant distortions of the octahedra. Additionally, based on electronic structure calculations, we find that the APB, which are negatively charged relative to the bulk, are likely to attract positively charged defects with high mobility. Overall, this study provides insights into the interaction between APB and ferroelectric domain walls in h-YbFeO$_3$ and highlights the need for defect engineering to improve the ferroelectric switching characteristics in these materials.

**Acknowledgements**

This work was supported by the National Science Foundation (NSF) under grant numbers DMR-2122070 (G.R.), and 2145797 (P.O.). Y.Y, X.L, and X.X. acknowledge the support by the National Science Foundation (NSF), Division of Materials Research (DMR) under Grant No. DMR-1454618 and by the Nebraska Center for Energy Sciences Research (NCESR). The Microscopy work was conducted as part of a user project at the Center for Nanophase Materials Sciences (CNMS), which is a US Department of Energy, Office of Science User Facility at Oak Ridge National Laboratory. This work used computational resources through allocation DMR160007 from the Advanced Cyberinfrastructure Coordination Ecosystem: Services & Support (ACCESS) program, which is supported by NSF grants # 2138259, #2138286, #2138307, #2137603, and #2138296. R.M. is grateful for the mentorship and the opportunities he received while working as a postdoctoral researcher with Steve Pennycook.## Data Availability

Data presented in this article is openly available in Zenodo at https://doi.org/10.5281/zenodo.13358523.

Page 14 of 16

**REFERENCES**

1. Sugiyama I, Shibata N, Wang Z, et al. 2013 Nature Nanotechnology 8(4):266-70
2. Ikuhara Y 2009 Progress in Materials Science 54(6):770-91
3. Catalan G, Seidel J, Ramesh R, Scott JF 2012 Reviews of Modern Physics 84(1):119-56
4. Li M, Huang Z, Tang C, et al. 2019 Advanced Functional Materials 29(49):1906655
5. Rojac T, Bencan A, Drazic G, et al. 2017 Nature Materials 16(3):322-7
6. Kim YM, Morozovska A, Eliseev E, et al. 2014 Nature Materials 13(11):1019-25
7. Kalinin SV, Rodriguez BJ, Borisevich AY, et al. 2010 Advanced Materials 22(3):314-22
8. Farokhipoor S, Magen C, Venkatesan S, et al. 2014 Nature 515(7527):379-83
9. Meier D, Selbach SM. 2021 Nature Reviews Materials 7(3):157-73
10. Ishikawa R, Okunishi E, Sawada H, et al. 2011 Nat Mater 10(4):278-81
11. Krivanek OL, Chisholm MF, Nicolosi V, et al. 2010 Nature 464(7288):571-4
12. Nellist PD, Chisholm MF, Dellby N, et al. 2004 Science 305(5691):1741
13. Pennycook SJ, Jesson DE 1991 Ultramicroscopy 37(1):14-38
14. Pennycook SJ, Jesson DE 1990 Phys Rev Lett 64(8):938-41
15. Browning ND, Chisholm MF, Pennycook SJ 1993 Nature 366(6451):143-6
16. Oxley MP, Lupini AR, Pennycook SJ 2017 Rep Prog Phys 80(2):026101
17. Pennycook SJ 2015 MRS Bulletin 40(01):71-8
18. Gazquez J, Guzman R, Mishra R, et al. 2016 Advanced Science 3(6):1500295:1-8
19. Biškup N, Salafranca J, Mehta V, et al. 2014 Phys Rev Lett 112(8)
20. Klie RF, Buban JP, Varela M, et al. 2005 Nature 435(7041):475-8
21. Kim M, Duscher G, Browning ND, et al. 2001 Phys Rev Lett 86(18):4056-9
22. Yan Y, Chisholm MF, Duscher G, et al. 1998 Phys Rev Lett 81(17):3675-8
23. Pennycook SJ, Nellist PD 2011 Springer Science & Business Media
24. Kimura T, Goto T, Shintani H, et al. 2003 Nature 426(6962):55-8
25. Wang W, Zhao J, Wang W, et al. 2013 Phys Rev Lett. 110(23):237601
26. Sinha K, Wang H, Wang X, Zhou L, Yin Y, Wang W, et al. 2018 Phys Rev Lett. 121(23):237203
27. Yen F, Dela Cruz C, Lorenz B, et al. 2007 Journal of materials research 22(8):2163-73
28. Jeong YK, Lee J-H, Ahn S-J, et al. 2012 J Am Chem Soc 134(3):1450-3
29. Xu X, Wang W 2014 Modern Physics Letters B 28(21):1430008
30. Lilienblum M, Lottermoser T, Manz S, et al. 2015 Nature Physics 11(12):1070-3
31. Mundy JA, Brooks CM, Holtz ME, et al. 2016 Nature 537(7621):523-7
32. Li X, Yun Y, Thind AS, et al. 2023 Sci Rep 13(1):1755
33. Barrozo P, Smabraten DR, Tang YL, et al. 2020 Adv Mater 32(23):e2000508
34. Skjaervo SH, Wefring ET, Nesdal SK, et al. 2016 Nat Commun 7:13745
35. Evans DM, Småbråten DR, Holstad TS, et al. 2021 Nano Letters 21(8):3386-92
36. Gelard I, Jehanathan N, Roussel H, et al. 2011 Chemistry of Materials 23(5):1232-8
37. Baghizadeh A, Vieira JM, Gonçalves JN, et al. 2016 The Journal of Physical Chemistry C 120(38):21897-904
38. Deng S, Cheng S, Liu M, et al. 2016 ACS Applied Materials & Interfaces 8(38):25379-85
39. Yun Y, Buragohain P, Thind AS, et al. 2022 Physical Review Applied 18(3):034071
40. Allen LJ, Findlay S 2015 Ultramicroscopy 151:11-22
41. Blöchl PE 1994 Physical review B 50(24):17953
42. Kresse G, Furthmuller J 1996 Physical Review B 54(16):11169-86
43. Perdew JP, Ruzsinszky A, Csonka GI, et al. 2008 Phys Rev Lett 100(13):136406
44. Dudarev SL, Botton GA, Savrasov SY, et al. 1998 Physical Review B 57(3):1505-9
45. Griffin SM, Reidulff M, Selbach SM, et al. 2017 Chemistry of Materials 29(6):2425-34

46. Liu J, Sun TL, Liu XQ, et al. 2018 Advanced Functional Materials 28(13):1706062
47. Giraldo M, Meier QN, Bortis A, et al. 2021 Nature communications 12(1):3093
48. Kumagai Y, Spaldin NA 2013 Nature communications 4(1):1540
49. Artyukhin S, Delaney KT, Spaldin NA, et al. 2014 Nature materials 13(1):42-9
50. Skjærvø SH, Meier QN, Feygenson M, et al. 2019 Physical Review X 9(3):031001
51. Huang F-T, Wang X, Griffin SM, et al. 2014 Phys Rev Lett 113(26):267602
52. Holtz ME, Shapovalov K, Mundy JA, et al. 2017 Nano Lett 17(10):5883-90
53. Meier QN, Lilienblum M, Griffin SM, et al. 2017 Physical Review X 7(4):041014
54. Matsumoto T, Ishikawa R, Tohei T, et al. 2013 Nano Lett 13(10):4594-601
55. Gao P, Nelson CT, Jokisaari JR, et al. 2011 Nature communications 2(1):591
56. Zhang D, Sando D, Sharma P, et al. 2020 Nature communications 11(1):349